# Characterizing the temporally stable structure of community evolution in intra-urban origin-destination networks

Xiao-Jian Chen[a,b,+], Yuhui Zhao[c,d,+], Chaogui Kang[e], Xiaoyue Xing[a,b], Quanhua Dong[a,b], Yu Liu[a,b,*]

[a]Institute of Remote Sensing and Geographical Information Systems, School of Earth and Space Sciences, Peking University, Beijing, China;

[b]Beijing Key Lab of Spatial Information Integration & Its Applications, Peking University, Beijing, China; [c]Research Center for Intelligent Society and Governance, Zhejiang Lab, Zhejiang, China;

[d]Zhejiang Pilot Laboratory of Philosophy and Social Science, Zhejiang, China;

[e]National Engineering Research Center of Geographic Information System, China University of Geosciences, Wuhan, China

[+]Both authors contribute equally to this work
*Corresponding author at: Peking University, Beijing 100871, China
E-mail addresses: cxiaojian@pku.edu.cn (X.J. Chen), zhaoyuhui@whu.edu.cn (Y. Zhao), kangchaogui@cug.edu.cn (C. Kang), xyxing@pku.edu.cn (X. Xing), dqh@pku.edu.cn (Q. Dong), liuyu@urban.pku.edu.cn (Y. Liu)

## Abstract

Intra-urban origin-destination (OD) network communities evolve throughout the day, indicating changing groups of closely connected regions. Under this variation, groups of regions with high consistency of community affiliation characterize the temporally stable structure of the evolution process, aiding in comprehending urban dynamics. However, how to quantify this consistency and identify these groups are open questions. In this study, we introduce the consensus OD network to quantify the consistency of community affiliation among regions. Furthermore, the temporally stable community decomposition method is proposed to identify groups of regions with high internal and low external consistency (named "stable groups"), where each group consists of temporally stable cores and attaching peripheries. Wuhan taxi data is used to verify our methods. On the hourly time scale, eleven stable groups containing 82.9% of regions are identified. This high percentage suggests that dynamic communities can be well organized via cores. Moreover, stable groups are spatially closed and more likely to distribute within a single district and separated by water bodies. Cores exhibit higher POI entropy and more healthcare and shopping services than peripheries. Our methods and empirical findings contribute to some practical issues, such as urban area division, polycentric evaluation and construction, and infectious disease control.

## 1. Introduction

Intra-urban movements reflect the intensity of connection between regions, which is characterized as the origin-destination (OD) network (X. Liu et al., 2015). To reveal the spatial organization, the urban OD network is often partitioned into multiple groups called "communities". Communities, originally proposed in the field of complex networks, refer to groups of regions (i.e., the spatial units of urban space) that have denser connections within the group and sparser connections between different groups (Mohamed et al., 2019). Therefore, communities are mesoscale representations of urban structure, and enable us to redefine urban geographical borders (Rinzivillo et al., 2012; Sobolevsky et al., 2013), design transportation systems (Yildirimoglu & Kim, 2018), or identify urban transformation (Zhong et al., 2014), to name a few.

Communities are constantly evolving throughout the day (M. Zhou et al., 2016). These variations are attributed to the differences in land functions of each region (Jia et al., 2022). For instance, in the morning, communities form between residential and working areas. During working hours, densely populated working areas become highly connected to support economic activities. However, at night, mixed behaviors like going home and entertaining form complex regional connections (X. Chen et al., 2022). Therefore, the closely connected regions are diurnal change.

Under this change, are there any regions that tend to belong to the same community? Groups of regions with high consistency of community affiliation suggest relatively frequent movements internally throughout the day, forming self-contained local spaces that characterize temporally stable structures in urban dynamics. Previous works mainly focus on community changes, exploring the evolution pattern from this stability perspective is still limited. How to quantify the consistency of community affiliation and identify groups of high-consistency regions are open questions. To fill these gaps, the main contributes are from the following three perspectives:

- The consensus OD network is introduced to describe the consistency of community affiliation, with the weighted edge representing consistency between two regions. This is inspired by the concept of consensus network (Lancichinetti & Fortunato, 2012), which was originally proposed for measuring the stability of community partitioning results.
- The temporally stable community decomposition (TSCD) method is proposed to identify a group structure of high-consistency regions. This structure comprises multiple groups, named "stable groups", each with high consistency within and low consistency between groups. A stable group is comprised of temporally stable cores attaching peripheries (see a toy example in Fig. 1).
- The spatial distributions of stable groups and the urban environmental factors in cores and peripheries are explored through map visualization analysis and logistic regression, respectively.

The Wuhan taxi dataset is used as a case study. The result shows that 82.9% of the regions are characterized by the group structure. This suggests that the daily variation of communities can be well represented as multiple groups organized via temporally stable cores. Furthermore, these groups are spatially closed and more likely to distribute within a single district and separated by water bodies (e.g., rivers and lakes) (Fig. 1). Compared to peripheries, cores have a higher point of interest (POI) entropy and more shopping and healthcare services.

Our results show several potential contributions in practical applications. One is to guide the delineation of urban management zones. Stable groups synthesize the variability of communities within a day, thus giving a more reasonable urban area division that captures the people's daily activities. Another contribution is the ability to evaluate the urban poly-centricity. Each group's cores can be considered as a subcenter. As such, our method can aid in identifying and assessing the city's centers based on temporal stability. Third, it contributes to developing more efficient policies in controlling infectious diseases. The mobility plays a pivotal role in the dissemination of epidemics (Alessandretti, 2022). In the early stage of an outbreak, monitoring and devising intervention strategies based on stable groups and focusing on targeting transmission between cores can help effectively prevent the rapid propagation of the epidemic.

This paper is organized as follows. Section 2 introduces the related work about the evolution of communities. Section 3 presents the data and methodology of the framework. Section 4 provides empirical analysis results. In section 5, the origins of stability are discussed. Following with the conclusion and future work in Section 6.

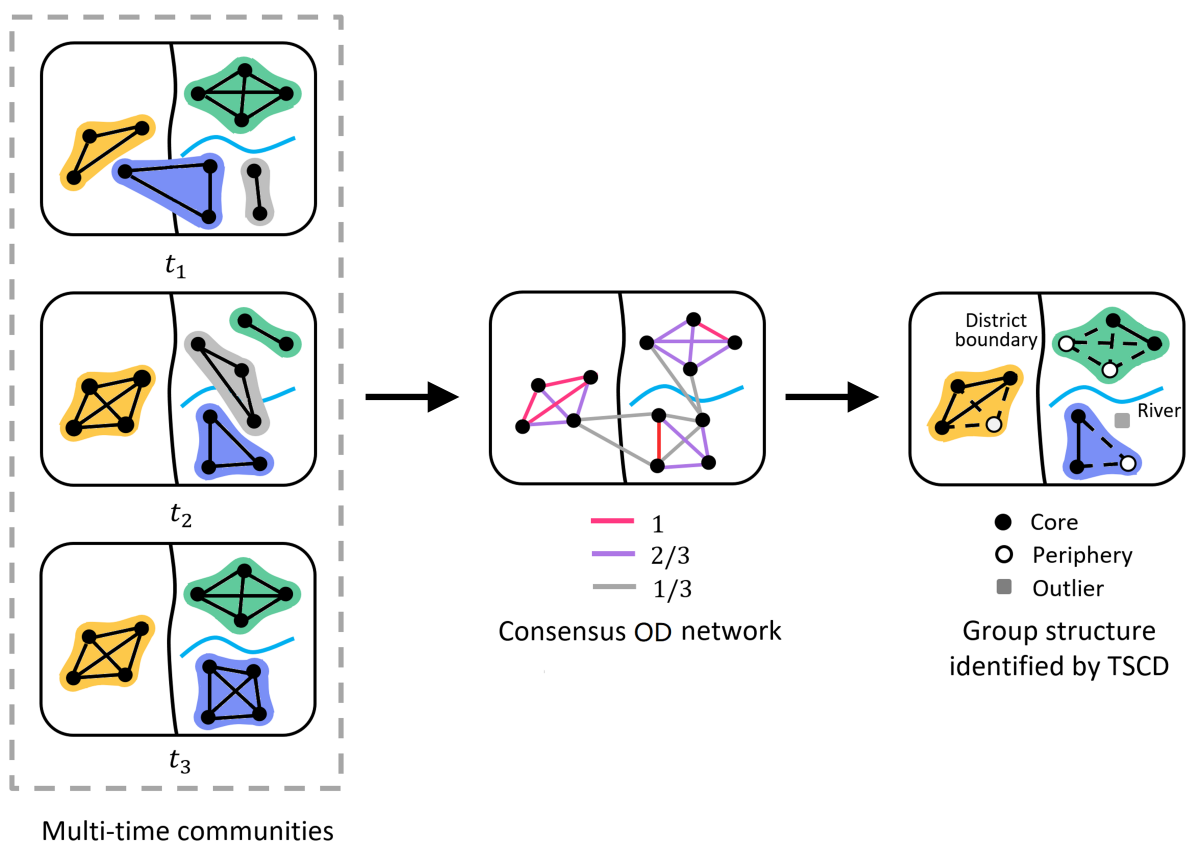

Fig. 1. A toy example of the consensus OD network and the group structure identified by the TSCD method.

## 2. Related work

### 2.1. Evolution of communities

The communities of the intra-urban OD network represent the urban organizational structure by identifying strongly connecting spatial blocks (Bakillah et al., 2015). Communities vary at different periods due to the natural temporal rhythm of movements. This evolution reflects urban structural dynamics and has received much attention in recent years.

Some works intuitively showed the changes in the community at different periods from the perspective of visual analysis. For example, Walsh & Pozdnoukhov (2011) used spatial distribution and alluvial timeline diagrams to present the communities' changes during the day. Zhang & Ng (2021) visually showed the changes in the inclusion relationship between the community and the rich club during the peak and off-peak periods.

In addition to visual analysis, there were also quantitative studies on the community evolution. For instance, Zhou et al. (2016) compared various characteristics, such as code length, modularity, normalized mutual information, "split-join" distance, etc. The results showed that the fluctuations in community partitions were most significant during the peak and pre-/post-peak hours. Similarly, Song et al. (2021) discovered that the average network degree and community size of the top 15 communities varied throughout the day, reaching their peak values during morning and evening rush hours. Additionally, these values were significantly higher on weekdays compared to weekends. Liu et al. (2022) used the Jaccard index to match communities among different periods and studied the changes in various communities using seven evolution indicators including continuation, splitting, merging, growth, contraction, birth, and death proposed by Palla et al. (2007). Jia et al. (2022) proposed a spatiotemporal community detection method based on a multilayer network model and studied the occurrence, expansion, stability, shrinkage, and disappearance of spatiotemporal communities according to the community size over time.

Therefore, urban community changes were complex, manifested not only in the daily variation of community-based characteristics but also in the multiple evolutionary dynamics of division, expansion, and contraction across different communities. The high variability in urban community evolution makes it challenging for researchers to summarize patterns. These motivate us to examine this evolution from an invariance perspective to supplement the understanding.

## 2.2. Stability of community evolution

While not their primary focus, some of the above studies have more or less hinted at the existence of stability of evolution. This further indicates the existence of regions that tend to belong to the same community throughout the day.

The typical one was the work of Zhou et al. (2016). They found that despite significant changes during peak and pre-/post-peak hours, the community partitions show high similarity across different periods. The normalized mutual information values between any two periods were generally high, ranging from 0.8 to 1, indicating nearly identical communities. They attributed this to urban zoning and planning schemes that partially limited people's activities. In addition to the highly similar overall community structure, some communities in the suburbs were also directly observed to remain stable.

Also, Zhang & Ng (2021) and Jia et al. (2022) both observed some communities remained roughly stable. Chen et al. (2022) found few interactions across regions and rivers in recurrent communities at different times. Moreover, these four works, along with Walsh & Pozdnoukhov (2011) and Liu et al. (2022), plotted the specific spatial distribution of communities at different periods in their paper. We observed that there was a large regional overlap between communities at different times, which indicated the existence of high-consistency regions.

Therefore, there was a certain stability in the evolution of communities. This reflected in the high similarity of community partitions over time and the presence of some roughly unchanging communities. By visually checking the dynamic communities, the stability was speculated result of the city's district planning and the river's influence. However, research on the stability is still scattered and limited. A precise depiction of the consistency of community affiliation among regions, along with identifying the groups with high-consistency regions are required. Also, the relationships between the stability and urban environment are necessary to be further verified and explored.

## 3. Data and methodology

This section presents the data and methodological framework proposed in this study. Section 3.1 describe the data. Section 3.2 introduces the consensus OD network to quantify the consistency of community affiliation among regions. In Section 3.3, the TSCD method is proposed to identify a group structure of high-consistency regions, consisting of temporally stable core identification and periphery matching. Finally, logistic regression is described in Section 3.4 to explore the difference of urban environmental factors in the core and periphery.

### 3.1. Data description

The case study is in the main urban area of Wuhan, China, located between longitude 114.10º-114.42º (around 31 km) and latitude 30.45º-30.70º (around 28 km) with 11 districts in this study area (Fig. 2). The city is divided into east and west banks by the Yangtze River, which are connected by bridges. There are also a lot of water bodies (the blue patches) in Wuhan. For the OD network analysis, we use the most common 1 km×1 km grid as the spatial basic unit (Nanni et al., 2020), dividing the city into 31×28 regions.

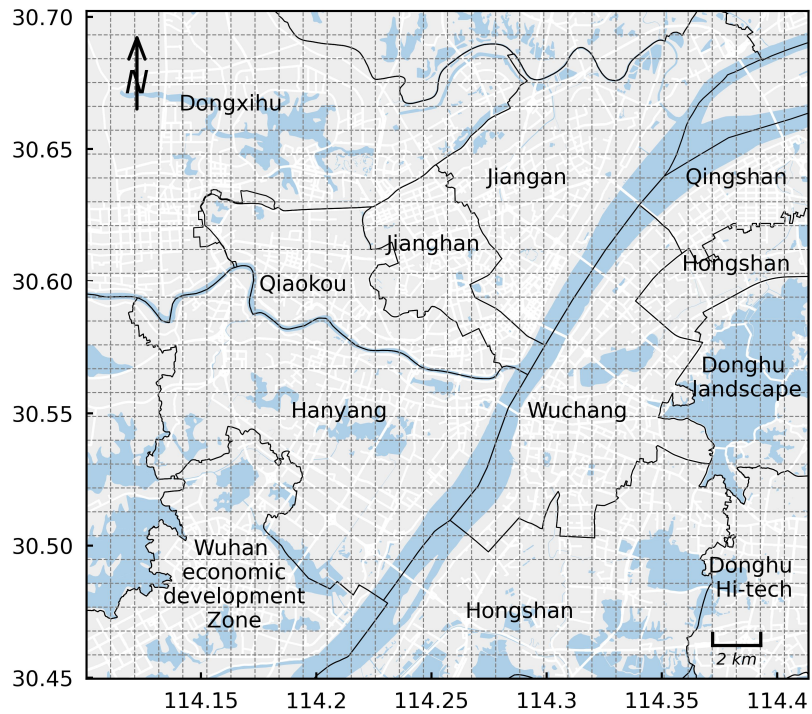

Fig. 2. The map of the main urban area of Wuhan.

Our study uses a taxi trajectory dataset from February 1 to August 10, 2015. As one of the most important means of public transportation in the city, taxis' trajectories are widely used as a proxy to describe the interactions between regions (W. Chen et al., 2022; Kang et al., 2022; Tang et al., 2015; Yu et al., 2023). GPS records with missing data due to weather conditions are removed based on the taxi company's recommendation, resulting in 35,549,882 trip records covering 173 days. Each trip includes accurate pick-up and drop-off coordinates and time, enabling the construction of OD networks on a 1 km × 1 km grid.

We use road networks, buildings, and POI data to describe the urban environment. The road networks and buildings data were obtained from the Wuhan Municipal Government in 2016, while the POI data was obtained from Amap, Inc. China in the same year. Despite its age, we mainly focus on the methods to characterize the temporally stable structure of communities using this data, which still allows us to validate the effectiveness of our methods. Finally, a total of thirteen types of POI data are considered, as displayed in Table 1.

Table 1. POI data

| POI data type | Primary information | Number |
| --- | --- | --- |
| 1. food | restaurant, pub, café, etc. | 55226 |
| 2. tourist attraction | memorial hall, national-level scenic spot, church, etc. | 1190 |
| 3. public facility | public toilet, newsstand, emergency shelter, etc. | 2373 |
| 4. enterprise | office buildings | 38983 |
| 5. shopping | store, mall, supermarket, etc. | 104460 |
| 6. transportation facility | parking lot, bus station, subway station, etc. | 17178 |
| 7. education | research institution, university, school, etc. | 16720 |
| 8. residential area | house, dormitory, villa, etc. | 13619 |
| 9. life service | hair salon, repair shop, dry cleaner, etc. | 55539 |

| 10. recreation | cinema, gym, game room, etc. | 8692 |
| 11. healthcare | hospital, pharmacy, clinic, etc. | 9534 |
| 12. government | governmental agency, police station, procuratorate, etc. | 10035 |
| 13. residential service | hotel, inn, apartment, etc. | 7049 |

## 3.2. Method for constructing consensus OD network

The consensus network concept was introduced by Lancichinetti & Fortunato (2012). In this network, edges are undirected and weighted, indicating the proportion of nodes belonging to the same community. Hence, we utilize the consensus network for multi-time OD networks, enabling us to assess the consistency between different regions

First, to describe the mobility within each period, multi-time OD networks are generated. According to a pre-defined time interval $\gamma$ (hour), a day is equally divided into $\tau = 24/\gamma$ segments: $\{[t_i, t_{i+1})|1 \leq i \leq \tau, t_1 < \cdots < t_{\tau+1}\}$. The multi-time OD networks are directed weighted networks defined as $N = \{N_1, \cdots, N_\tau\}$, where $N_i = (V_{N_i}, E_{N_i}, W_{N_i})$ represents the OD network. $V_{N_i}$, $E_{N_i}$, and $W_{N_i}$ are node, edge and weight sets, respectively. For the specific meanings, $V_{N_i}$ represents spatial units, $E_{N_i}$ represents movements from the origin to the destination, and $W_{N_i}$ represents the number of movements begin within the corresponding interval.

Second, Infomap (Rosvall & Bergstrom, 2008), a popular and well-performing algorithm designed for directed weighted network, is adopted for community detection in $N_i = (V_{N_i}, E_{N_i}, W_{N_i})$. Infomap converted the network community detection into a coding problem, and measured the quality of clustering by using the random walks to estimate the code length of the path in the graph. The idea was that a good group division resulted in a shorter code length. Therefore, it tried to find the group partition with the shortest random walk length through multiple iterations of optimization. As such, we have the multi-time community structures $C = \{C_1, \ldots, C_\tau\}$, where $C_i = \{c_1^i, \ldots, c_{r_i}^i\}$ is the partition result from $N_i = (V_{N_i}, E_{N_i}, W_{N_i})$.

Third, calculate the proportion of two nodes belonging to the same community, and construct the consensus OD network. Specifically, given the multi-time community structures $C = \{C_1, \ldots, C_\tau\} = \{\{c_1^i, \ldots, c_{l_i}^i\}\}, 1 \leq i \leq \tau\}$, the consensus OD network is an undirected weighted network defined as $G^c = (V^c, E^c, W^c)$, where $V^c = \bigcup_{1 \leq i \leq \tau} \bigcup_{1 \leq j \leq l_i} c_j^i$ is the union of all the nodes, the edge $e_{ij}^c \in E^c$ with weight $w_{ij}^c \in W^c$ represents the ratio of $v_i$ and $v_j$ belonging to the same community:

$$w_{ij}^c = \frac{\sum_{C_i \in C} \sum_{c_j^i \in C_i} \chi\{v_i \cup v_j \subseteq c_j^i\}}{\tau} \quad (1)$$

As such, $w_{ij}^c \in (0,1]$ and $w_{ij}^c = 1$ means $v_i$ and $v_j$ belong to the same community all the time (See a toy example in Fig. 1).

## 3.3. Temporally stable community decomposition

Based on the consensus OD network, TSCD is a two-step method that contains temporally stable core identification and periphery matching.

### 3.3.1. Temporally stable core identification

The most stable nodes in the dynamic communities are identified as temporally stable cores. To do this, we select the sub-network $G^{core}$ composed of edges with $w_{ij}^c = 1$ in the consensus network. Therefore, $G^{core}$ indicates nodes always belonging to the same community throughout the day. According to the definition of consensus network, $G^{core}$ has an interesting property, which we call "clique separability" (clique refers to a subgraph having connections between any two nodes). We use this property to extract multiple groups of temporally stable cores.

Specifically, clique separability means $G^{core}$ is composed of multiple cliques, and there is no connection between these cliques. This is due to connection transitivity between nodes. If $v_1$ and $v_2$ are connected, it means that they always belong to the same community. The situation is the same if $v_3$ connects $v_1$. Thus, $v_3$ also always belongs to the same community as $v_2$, and they are connected. To sum up, any two nodes are either disconnected or form a clique with other connected nodes, which proves the property.

Thus, temporally stable cores are initially defined as the cliques of $G^{core}$, represented by $D^{core} = \{T_1, \ldots, T_k\}$. However, it is necessary to merge some of them. This is because the uncertainty of the algorithm results or small changes in data can affect the community segmentation. This would further result in some cores that may not consistently belong to the same community, but often affiliate to the same one. Merging these cores prevents excessive fragmentation of group structure.

The merging procedure is conducted by reducing the threshold condition: if $T_i$ and $T_j$ form a clique through the consensus edges with $w_{ij}^c \geq \alpha$, then they are considered to be merged. Based on this pairwise merge operation, we further use three specific operations to complete multiple merges to get final groups of cores, recorded as $D_\alpha^{core} = \{T_{\alpha,1}^{merge}, \ldots, T_{\alpha,m}^{merge}\}$:

- Start merging $T_i$ with the smallest size (i.e., number of nodes, recorded as $S_{T_i}$) and only $T_j$ with $S_{T_j} \geq S_{T_i}$ can become the merged object.
- If $T_i$ can be merged into different $T_j$s, select the one with the largest size.
- For a merged group containing multiple core groups $\{T_{i_k}, k > 1\}$, the one with maximum size is the group representation. If a group representation is judged to be merged into another group representation, then two merged groups are merged together.

Users can set $\alpha$ according to research needs or experience. However, to assist with threshold setting, a two-stage heuristic algorithm is also provided. The key idea is gradually reducing $\alpha$ until all scattered small cores are merged. Specifically, it needs to predefine a size parameter (denoted as s) to indicate how small is considered small, such as $s = 5$. In the first stage, alpha just starts to decrease, and the size of the newly merged core groups (denoted as $s_{new}$) is allowed to be larger than s. When $s_{new} \leq s$ for the first time, the second stage is entered. The algorithm is considered to start to incorporate core groups with small sizes. $\alpha$ continues to decrease until it stops when $s_{new} > s$ again. The first stage is to merge as many cores as possible that do not belong to the same community by chance. The second stage is to further merge those scattered small core groups.

Noticing that the above heuristic method is not fully automated. Users need to determine the final results based on the size threshold $s$. Additionally, to ensure the merged cores belong to the same community for the majority of the time, $\alpha$ is recommended to be at least greater than 0.6. However, this two-stage heuristic method is still meaningful. It examines the results of different $\alpha$ to avoid fragmented group structure, making it a preferable alternative to directly setting $\alpha$. Section 4.2.1 demonstrates that different $s$ can yield the same result, highlighting the robustness of $s$ in our case. The development of an automated method is an important future direction.

### 3.3.2. Periphery matching

Then we need to match other nodes to the stable cores $\{T_1^{merge}, \ldots, T_m^{merge}\}$. We assess the likelihood of nodes belonging to the same community as stable cores by the following matching vector:

$$V_i^{match} = (v_{i,1}^{match}, \ldots, v_{i,m}^{match}, v_{i,m+1}^{match}) \quad (2)$$

where $v_{i,k}^{match}, 1 \leq k \leq m$ indicates the number of times that $v_i$ and the nodes in $T_k^{merge}$ belong to the same community, $v_{i,m+1}^{match}$ indicates the number of times that $v_i$ does not belong to the same community as any nodes in $T_k^{merge}, 1 \leq k \leq m$. If $\max_j v_{i,j}^{match} \geq v_{i,m+1}^{match}$, $v_i$ is matched to $\operatorname{argmax}_j v_{i,j}^{match}$ and called periphery, else $v_i$ is called an outlier.

### 3.4. Logistic regression

For cores and peripheries, the following features are calculated in the corresponding region: the length of the roads, the area of buildings, the number of thirteen kinds of POIs, and the mixing degree of POIs. The mixing degree of POI in $v_i$ is calculated by the entropy value (Brown et al., 2009):

$$entropy_i = \sum_{1 \leq j \leq 13} p_{ij} \ln(p_{ij}) \quad (3)$$

where $p_{ij}$ refers to the proportion of POIs of the category $j$ in $v_i$, and has $\sum_{1 \leq j \leq 13} p_{ij} = 1$, for any $i$.

Then, features with strong collinearity are removed. This is conducted by iteratively deleting the feature with the largest variance inflation factor (VIF) (Craney & Surles, 2002) until all VIF<5, where 5 is recommended as a strict threshold to exclude collinearity (Marcoulides & Raykov, 2019). Finally, the logistic regression (Sperandei, 2014) is conducted by these features to classify the core and periphery. The features' coefficients are analyzed to reveal the influence of urban environmental factors.

# 4. Results

## 4.1. Construction of consensus OD network

The temporal resolution in this study is set to one hour, which is commonly used for capturing the evolution pattern of urban OD networks by many studies (Jia et al., 2022; M. Zhou et al., 2016). As such, a day is divided into 24 periods. The movements of passengers are classified into each period by the pick-up time. For each OD network, higher-weighted edges covering fifty percent of the cumulative flows are retained to describe the primary interactions. Fig. 3 shows the 24-hour taxi volume change in Wuhan. The decrease in passenger flow during 16:00-18:00 is due to the shift time of taxi drivers in Wuhan, which affects taxi services. In Wuhan, it is common for two drivers to share a vehicle at different times to keep the taxis operating.

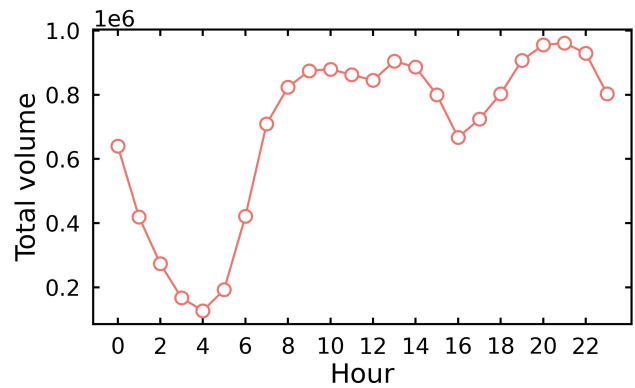

Fig. 3. The total volume of passengers' movements in each period. The hour in the x-axis indicates the period [hour-2, hour).

Fig. 4 shows the spatial distribution of communities at different periods. Communities are spatially closed, which aligns with findings from previous studies and can be attributed to the distance decay effects of spatial movements. Also, except for the two largest communities on the west and east banks of rivers, communities appear to be within a single district or not expanding much. Moreover, communities also have fewer connections across the Yangtze River and water bodies. These findings are similar to previous studies. Finally, although communities change from time to time, the community structures appear to have a high degree of similarity.

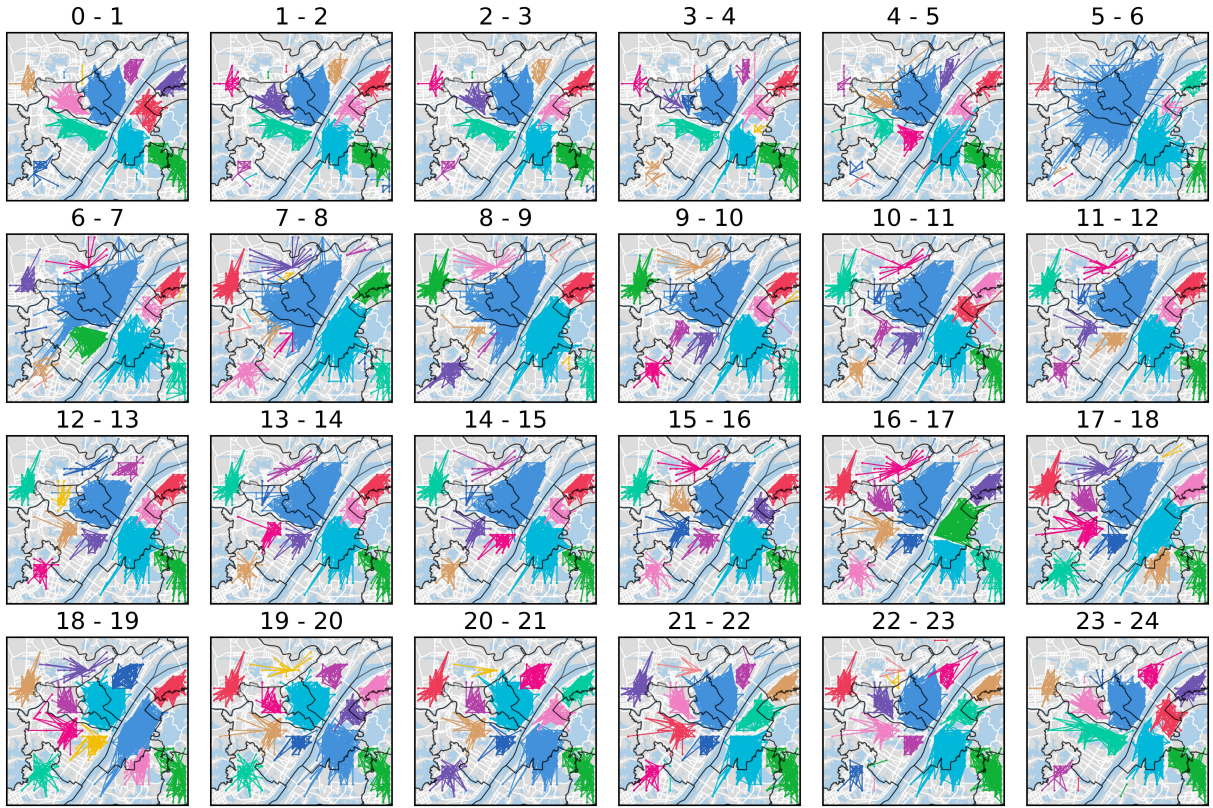

Fig. 4. Spatial distributions of communities in each period, where periods are shown in titles.

Fig. 5(a) shows the spatial distribution of consensus OD networks $G^C$. A large number of high-weight edges are observed, indicating the existence of a stable community structure. To test whether the observed weights are significantly high, we compare the result with a random procedure as a null model. We shuffle the observed community labels in each period and construct the corresponding random consensus OD network. Fig. 5(b) shows the spatial distribution of a random consensus OD network, showing that the random consensus OD network is composed of many lower-weighted edges connected across regions, rather than higher-weighted edges forming connections in relatively fixed blocks of regions.

To further verify the difference quantitatively, we compare the probability distribution of observed consensus weights $W^c$ with 1000 times random consensus weights $\{W_i^{c,random}, 1 \leq i \leq 1000\}$. The results show that no $W_i^{c,random}$ belong to the same probability distribution as $W^c$ by the Kolmogorov-Smirnov statistic (Hodges Jr, 1958). Fig. 5(c) shows the probability functions of $W^c$ and $\cup_{1 \leq i \leq 1000} W_i^{c,random}$. Despite the rapid decline of the random curve, the observed curve exhibits the characteristic of a long-tailed distribution and continues to the maximum value of 1. This significant difference further indicates that the consensus OD network has higher edge weights than the random ones, thus illustrating that some nodes tend to form a temporally stable community in OD networks.

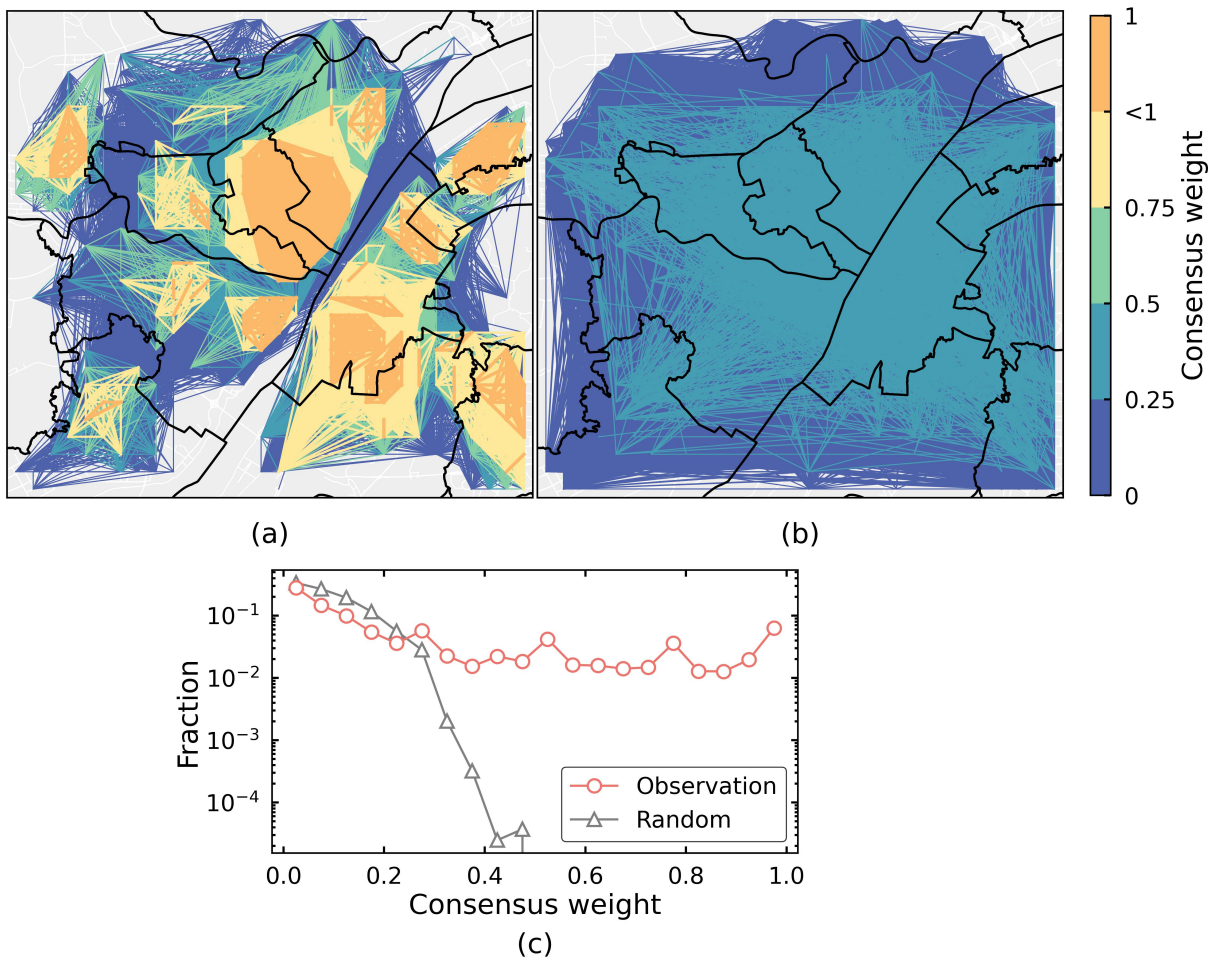

Fig. 5. The spatial distribution of the (a) observed consensus OD network and (b) a random one. (c) The fraction distribution of consensus weights for observation ($W^c$) and random ($\cup_{1 \leq i \leq 1000} W_i^{c,random}$).

## 4.2. Group structure identification by TSCD

### 4.2.1. Stable cores identification

Figure 6(a) shows the spatial distribution of $G^{core}$, with distinct colors representing disjoint cliques. It finds that some of these clusters are very close in space. Then, $G^{core}$ are merged using the proposed two-step heuristic algorithm. The size threshold is set as $s = 5$. As shown in Fig. 6(d), the first stage is $\alpha \geq 22/24$. Fig. 6(b) shows the spatial distribution of $G^{core}_{22/24}$, where the red circles highlight four typical merged core groups that are adjacent to each other in space. However, there are still some tiny core groups pointed by arrows in Fig. 6(b).

In the second stage when $\alpha \leq 21/24$, $\alpha$ is continuously reduced. By analyzing the changes of size (Fig. 6(d)), we determine the final merge threshold as $18/24 = 0.75$. This is because at this threshold, the maximum size for the new merge is 3. Continue decreasing $\alpha$ until 12/24 to trigger new merges. However, we won't choose $\alpha = 12/24$. This threshold implies that only 50% of the periods the cores belong to the same community, which doesn't necessarily mean they are together for most of the time. Moreover, the size of the new merge exceeds $s = 5$, suggesting that this threshold starts merging larger-sized cores.

Fig. 6(c) shows the final 11 core groups, where the trivial cores have been merged. In fact, setting $s$ as 3 to 5 yields the same result. Also, following the suggestion in Section 4.2.1, we only need to check the result when $\alpha > 0.6$ to determine the threshold.

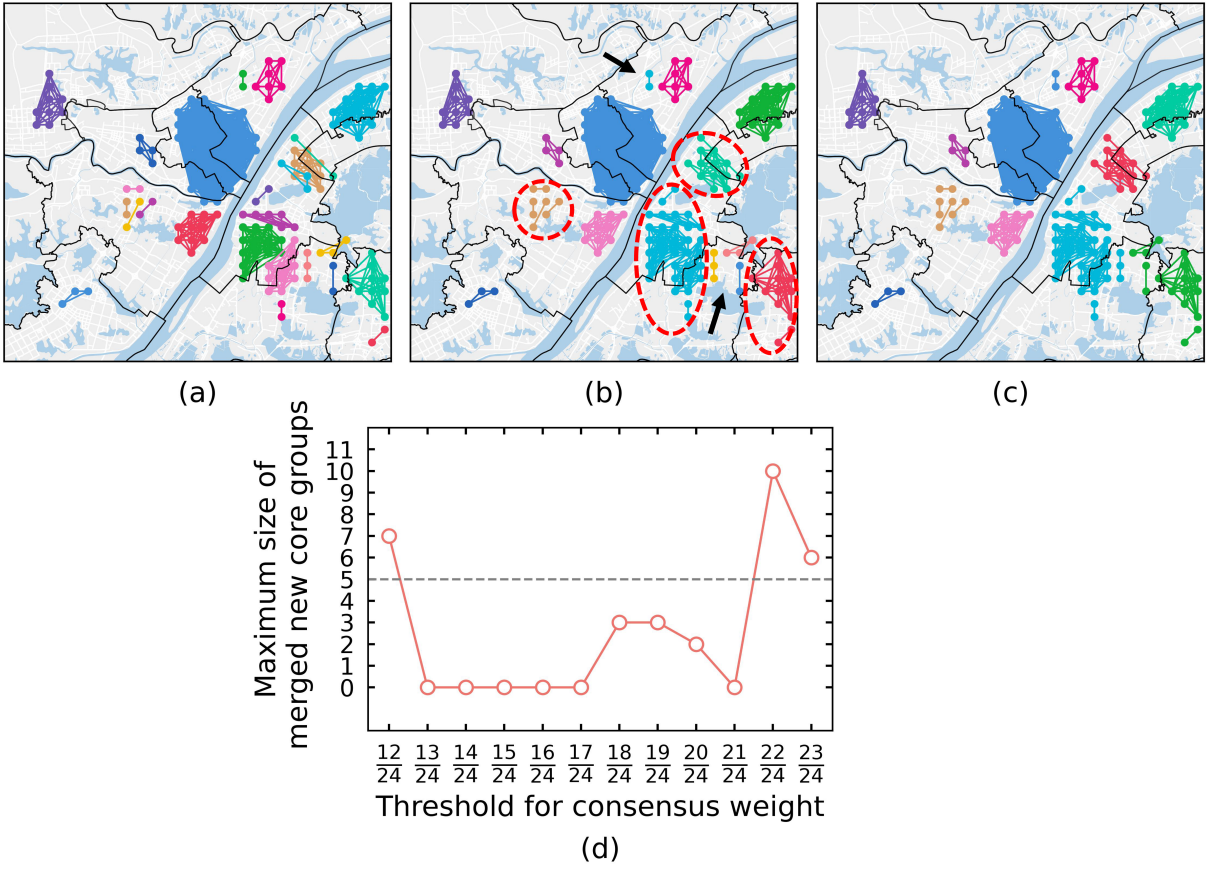

Fig. 6. The spatial distribution of (a) $G^{core}$, (b) $G^{core}_{22/24}$ and (c) $G^{core}_{18/24}$. (d) The maximum size of merged new core groups with different thresholds.

### 4.2.2. Periphery matching

After extracting stable cores, other regions are classified into these cores based on their matching vectors $V_i^{match} = (v_{i,1}^{match}, \ldots, v_{i,m}^{match}, v_{i,m+1}^{match})$. To determine if each region is more stably matched to the same core, we compared its occurrence in twenty-four periods to its largest component value in $V_i^{match}$ (Fig. 7(a)). The observations reveal that points are concentrated around $y = x$, indicating that regions tend to belong to the same core community when they appear.

The pair values of the 1st and 2nd largest components are mainly concentrated in the lower right, indicating that the 2nd one is significantly smaller than the 1st (Fig. 7(b)). This suggests a relatively unique matching. Therefore, peripheries, while not having the most stable community relationship, still form a relatively unique stable community with a core group.

Additionally, the occurrences of outliers in a day and to the number of non-core matches $v_{i,m+1}^{match}$ are similar (Fig. 7(c)), indicating a low frequency of outliers and cores within the same community.

Overall, out of 468 regions with OD interactions, there are 164 (35.0%) cores, 224 (47.9%)

periphery, and 80 (17.1%) outliers. This suggests that communities are dynamic, but can be well organized as multiple stable cores with attaching peripheries, except for a few outliers that do not form groups with stable cores.

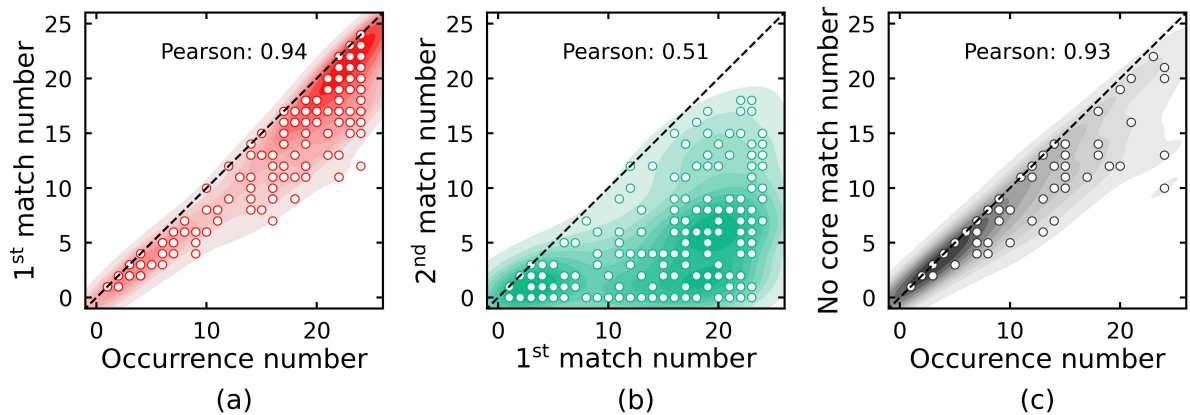

Fig. 7. (a) The occurrence number versus the 1$^{st}$ largest components of $V_i^{match}$ for peripheries. (b) The 1$^{st}$ largest components versus 2$^{nd}$ of $V_i^{match}$ for peripheries. (c) The occurrence number versus $v_{i,m+1}^{match}$ for outliers.

### 4.3. Influences of urban environment

This subsection aims to reveal the relationships between the stable structure and the urban environment. First, the spatial distributions of stable groups are analyzed through map visualization, combined with district borders and water bodies. We use logistic regression to explore urban environmental differences between cores and peripheries.

#### 4.3.1. Spatial distribution patterns

Fig. 8 shows the results of stable groups. It illustrates each group forming a spatially connected block, typically confined by water bodies and within a single district. The rough outline of each group, based on results from Fig. 8(d-n), is manually delineated on the map for clarity in Fig. 8(b).

On the one hand, groups tend to have fewer connections across water bodies. It finds that no groups cross the Yangtze River, and six inner-city water bodies (marked by blue circles in Fig. 8(b)) and the Hang River, limit community formation despite bridge or road access. For instance, the Hang River divides northern groups (1 and 10) and southern ones (8 and 6). Water bodies separate groups 2, 3, and 5, and also groups 6, 8, and 11. Many outlier regions exist in the Dongxihu district. Despite lacking stable cores, these regions likely belong to the same community. However, being flanked by two water bodies hinders these outliers from forming stable communities with the established cores of groups 7 and 9.

On the other hand, although communities have sprawled to a certain extent, a single community mainly distributes within a single district, such as groups 3-11. The certain alignment of water bodies and district boundaries could possibly contribute to this observed phenomenon. However, group distributions do have a certain correlation with district boundaries. For example, group 3 is mostly located in Donghu Hi-tech and does not merge with Group 2 despite large land connections. In the same vein, groups 7 and 10, as well as groups 5 and 4, remain separate within their specific districts. Similarly, outliers in the Dongxihu, despite proximity, don't form a stable community with the cores in group 1.

The two largest groups 1 and 2 span several districts, also located in the two centers of Wuhan (X. Chen et al., 2022). Group 1 spans Qiaokou, Jianghan, and Jianggan districts, formerly one district, "Hankou", a historic economic and financial hub. It later separated into three districts for easier management. Currently, group 1 is Wuhan's economic and financial core, housing its bustling CBD. Group 2 covers Wuchang and Hongshan. These two districts were together decades ago and were later separated. Both are Wuhan's cultural and political centers, with numerous schools and Wuhan government agencies.

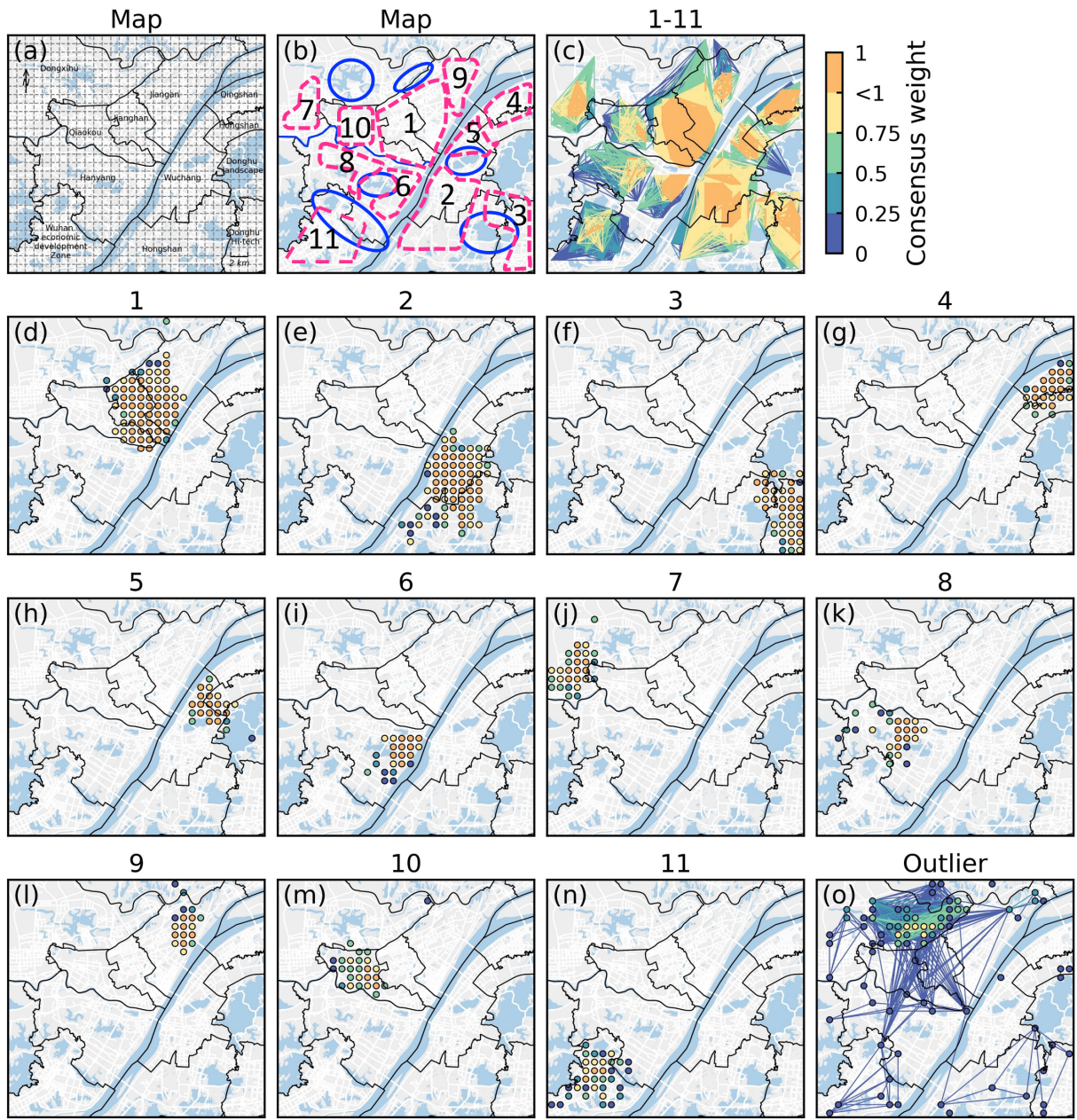

Fig. 8. (a) The map of the study area in Wuhan. (b) The outline of each group (shown in pink lines) and the water bodies between the groups (shown in blue lines). (c) The consensus OD network of the identified eleven stable groups. (d-n) The nodes of each group. (o) The consensus OD network of outliers. The color is marked by the maximum consensus weight of each node's edge. As such, the red ones indicate the cores.

## 4.4. Urban environmental factors of cores and peripheries

The differences of urban environmental factors in cores and peripheries are explored by logistics regression. In our case, two types of POIs: life service (VIF=12.46) and food (VIF=9.64) are deleted one by one, and the remaining features are shown in Table 2 (with their spatial distributions shown in Fig. 9).

Table 2. The statistics of urban environmental factors without collinearity.

| Feature | VIF | Min-max (mean) per grid |
| --- | --- | --- |
| tourist attraction | 1.08 | 0-97 (2.07) |
| public facility | 2.39 | 0-49 (5.20) |
| company | 2.54 | 0-990 (87.32) |
| shopping | 4.11 | 0-1937 (247.02) |
| transportation facility | 4.79 | 0-161 (38.94) |
| education and culture | 2.44 | 0-385 (37.37) |
| residential area | 1.78 | 0-445 (29.11) |
| recreation | 4.18 | 0-187 (20.36) |
| healthcare | 3.56 | 0-129 (23.05) |
| government | 3.10 | 0-159 (23.28) |
| residential service | 2.12 | 0-229 (16.91) |
| street length (m) | 1.58 | 0-15016.11 (5921.64) |
| building area ($m^2$) | 2.25 | 0-420503.69 (149272.60) |
| POI entropy | 1.87 | 0.78-2.28 (1.91) |

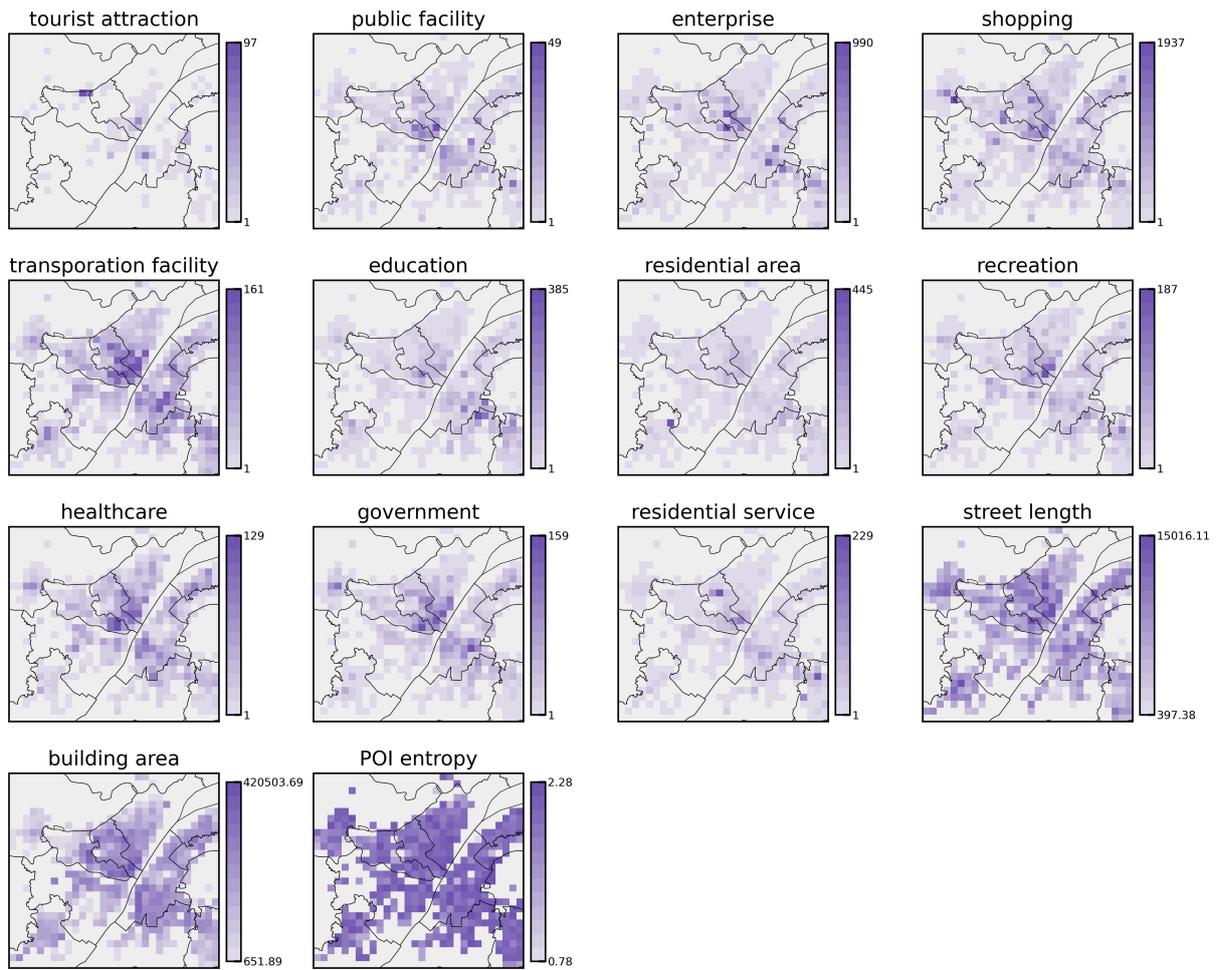

Fig. 9. The spatial distributions of each feature in cores and peripheries.

The result for logistic regression is shown in Table 3, where the positive class is core. The overall classification accuracy is 0.81 with 0.87 and 0.73 for the cores and peripheries. This indicates, in our case, that the differences in the urban environment of the cores and peripheries can be well captured by the features considered.

Significant urban characteristics distinguishing cores from peripheries include high POI entropy and a large number of shopping and healthcare services. Higher POI entropy indicates a better functional mix of a grid. High-mix grids together can meet people's varying daily travel needs, thereby forming stable cores.

Stable cores have more healthcare and shopping services than the peripheries, possibly due to urban planning. In Wuhan, healthcare entities like hospitals and pharmacies are primarily government-planned, mainly located in sub-centers with high visits of each district, thus enabling convenient access for residents. Subcenters, with their high functional mix (Yue et al., 2017), tend to easily form stable cores. A similar pattern is seen with shopping service. On the one hand, large shopping malls are mainly planned in sub-central areas with a lot of visits. On the other hand, small convenient stores, and supermarkets also tend to spontaneously agglomerate in places with high visits.

Table 3. The logistic regression for the cores and peripheries.

| Feature | Coef. | P-value |
| --- | --- | --- |
| tourist attraction | -0.0016 | 0.945 |
| public facility | 0.020 | 0.613 |
| enterprise | -0.0037 | 0.084 |
| **shopping** | **0.0028*** | **0.029** |
| transportation facility | 0.019 | 0.062 |
| education | -0.0076 | 0.181 |
| residential area | -0.0040 | 0.627 |
| recreation | 0.023 | 0.180 |
| **healthcare** | **0.035*** | **0.007** |
| government | -0.00080 | 0.937 |
| residential service | 0.010 | 0.329 |
| street length (meters) | 0.000037 | 0.545 |
| building area (meter$^2$) | 0.0000050 | 0.061 |
| **POI entropy** | **2.47*** | **0.035** |
| const. | -8.08** | 0.001 |

| Performance | |
| --- | --- |
| Type | Accuracy |
| core | 0.87 |
| periphery | 0.73 |
| all | 0.81 |

Core is the positive class. Note: ∗ p <0.05; ∗∗ p <0.01.

## 5. Discussion

In this study, we observed that some regions (i.e., temporarily stable cores) belong to the same community consistently throughout the day. It raises the question of why certain regions maintain such stability in dynamic networks. Combining previous studies as well as observations in this study, we hypothesize the following three reasons as the origin of stability.

First, movements between regions are subject to distance decay at any time of the day (Y. Liu et al., 2012). This means that spatially close regions tend to be densely connected, which makes it possible to form stable communities.

Second, the advocate of mix-use development in urban planning (Yue et al., 2017) may result in stable communities. Urban planners prefer high-mixed land use with jobs, residences, and entertainment. Because this self-containment planning promotes sustainable transport and urban development by reducing travel distance (X. Zhou et al., 2022). Highly mixed areas of spatial proximity meet daily travel needs and may support temporally stable communities. In our case, stable communities generally exist within a single district, verifying this hypothesis to some extent as the district is the basic scope of urban planning (Geneletti et al., 2020).

Third, water bodies promote the formation of stable communities. Although people can move across rivers or lakes by detours or bridges, the water bodies are still the barriers to movements (Kondolf & Pinto, 2017; Tóth et al., 2021). This makes urban planning also tend to build areas that meet mixed functions in areas surrounded by water bodies, thereby promoting the formation of stable community structures. Therefore, to meet the daily needs of people, urban planning often develops high-mixed-use constructions in areas surrounded by water bodies, thus promoting the formation of stable community structures.

## 6. Conclusion and future work

This study revealed the stable patterns of intra-urban community evolution within a day from the following three perspectives. First, the consensus OD network was introduced to quantify the consistency of community affiliation among regions, offering a node-level stability characteristic. Regions exhibited greater consistency compared to random results.

Second, TSCD was proposed to identify groups with high internal and low external consistency. Each group comprised temporally stable cores and attaching peripheries. Cores indicated regions with unchanged community relations, while attaching peripheries represented regions with high consistency with the cores of the group compared to other groups. As such, TSCD provided the natural representation of a stable structure. In our case, TSCD identified eleven groups, covering 82.9% of the regions (35.0% cores and 47.9% peripheries), suggesting that dynamic communities were well-organized by temporarily stable cores of multiple groups.

Finally, the spatial distributions and urban environmental factors were further analyzed for the above groups. In our case, it validated the previously observed effects of district borders and water bodies on stability. Furthermore, the temporally stable cores showed higher POI entropy and more healthcare and shopping services compared to the peripheries. These empirical findings also deepened our understanding of the relationship between intra-urban interactions and urban environments.

However, there are some limitations and work that need to be done in the future. First, more mobility data is required for the research. This paper mainly uses Wuhan's taxi data as an example to validate the proposed methods. Although widely used for analyzing regional spatial interaction patterns and community evolution (Jia et al., 2022), taxi data is biased and insufficient to fully capture the movement of people in cities. Additionally, comparing the stable structures across multiple cities to explore patterns and similarities/differences is an important research direction.

Another future direction is to analyze the spatio-temporal patterns of OD flows within the stable community structure. While our work primarily focuses on the relationship between stable structure and urban environments, further analysis of OD flows is necessary to deepen the characterization of stable structure. For instance, examining the variations in OD flows within individual groups and between groups.

Further improving the TSCD algorithm is also important in the future. Currently, cores require that community relations be completely consistent throughout the day, potentially overlooking sub-stable groups (such as the one in the Dongxihu district shown in Fig. 8(o)) and impeding the comprehensive understanding of the stable structure in OD evolution. Also, the automated method for determining the threshold value to merge cores is another critical work.